\documentclass[lettersize,journal]{IEEEtran}
\usepackage{amsmath,amsfonts}
\usepackage{algorithmic}
\usepackage{array}
\usepackage[caption=false,font=normalsize,labelfont=sf,textfont=sf]{subfig}
\usepackage{textcomp}
\usepackage{stfloats}
\usepackage{url}
\usepackage{verbatim}
\usepackage{graphicx}
\usepackage{color}
\def\BibTeX{{\rm B\kern-.05em{\sc i\kern-.025em b}\kern-.08em
    T\kern-.1667em\lower.7ex\hbox{E}\kern-.125emX}}
\usepackage{balance}
\begin{document}

\title{Analytic Network Traffic Prediction Based on User Behavior Modeling}

\author{Liangzhi Wang, Jiliang Zhang, Zitian Zhang, and Jie Zhang

\thanks{Liangzhi Wang is with the Department of Electronic and Electrical Engineering, the University of Sheffield, Sheffield, S10 2TN, UK. (e-mail: lwang85@sheffield.ac.uk).

Jiliang Zhang is with College of Information Science and Engineering, Northeastern University, Shenyang, China. (e-mail: zhangjiliang1@mail.neu.edu.cn).

Zitian Zhang is with School of Information and Electronic Engineering, Zhejiang Gongshang University, Hangzhou, China. (e-mail: zitian.zhang@mail.zjgsu.edu.cn).

Jie Zhang is with the Department of Electronic and Electrical Engineering, the University of Sheffield, Sheffield, S10 2TN, UK, and also with Ranplan Wireless Network Design Ltd., Cambridge, CB23 3UY, UK. (e-mail: jie.zhang@sheffield.ac.uk).}}

\markboth{Journal of \LaTeX\ Class Files,~Vol.~18, No.~9, September~2020}%
{How to Use the IEEEtran \LaTeX \ Templates}

\maketitle

\begin{abstract}

This paper proposes an interpretable user-behavior-based (UBB) network traffic prediction (NTP) method. Based on user behavior, a weekly traffic demand profile can be naturally sorted into three categories, i.e., weekday, Saturday, and Sunday. For each category, the traffic pattern is divided into three components which are mainly generated in three time periods, i.e., morning, afternoon, and evening. Each component is modeled as a normal-distributed signal. Numerical results indicate the UBB NTP method matches the practical wireless traffic demand very well. Compared with existing methods, the proposed UBB NTP method improves the computational efficiency and increases the predictive accuracy. 
\end{abstract}


\section{Introduction}
\IEEEPARstart {N}{owadays}, network traffic prediction (NTP), which analyzes historical traffic data to predict future traffic, is really crucial. 
With the rapid rise and ongoing expansion of Internet technology, users are increasingly voracious in their demand for network traffic. 
By 2025, there will be 5.7 billion Internet users worldwide and 32.1 billion connections, which will generate a vast amount of data, according to the prediction of Global System for Mobile communications Association (GSMA) [1].
Under the combined influence of massive data, heterogeneous networks, and terminal mobility, network efficiency is facing severe challenges [2].
The analysis and prediction technology for network traffic is regarded as one of the most significant methods to enhance network efficiency.
Currently, NTP has been widely researched as a time sequences forecasting problem [3].

The state-of-the-art regarding NTP mainly focuses on two directions, i.e., the statistics-based methods and the machine learning (ML)-based methods [4]. 
Among statistics-based methods, the Autoregressive Moving Average (ARMA) model and the Autoregressive Integrated Moving Average (ARIMA) model are the most commonly used models to extract linear features in historical data [2]. ARMA model is suitable for processing stationary time sequences [5], while the ARIMA model introduces a difference-stationary process, which enables the prediction of non-stationary time sequences [6-7]. For the nonlinear features, Anand \textit{et al.} [8] introduced the Generalized Auto-Regressive Conditional Heteroskedasticity (GARCH) model. However, traditional statistics-based methods do not perform very well in terms of prediction accuracy and computational efficiency.

ML-based methods represented by Deep Neural Networks (DNN) show superior performance in terms of prediction accuracy [9]. Multiple activation functions provide the capacity to understand complex nonlinear relationships. According to the comparison between ARIMA models and Long Short-Term Memory (LSTM) networks, the experiments in [9-10] demonstrated that LSTM has superior performance. However, hyper-parameter selection is complicated and difficult, and is related to efficiency and even accuracy. To improve efficiency of hyper-parameter selection, meta-learning scheme has been applied in the NTP field [11]. Zhang \textit{et al.} [3] introduced a deep meta-learning method to improve the learning ability and efficiency of DNN-based NTP models. ML-based methods construct the NTP as a black box. As a corollary, it has poor explanatory power. In addition, efficiency can be affected by the training time.

It is vital to note that traffic patterns are present in historical traffic data and, more importantly, that user behavior dominates the patterns.
The state-of-the-art is built on the former and overlooks the latter. 
They extract an aggregation of traffic patterns from historical data and store it with a specific model. The model is comprehensible for computer, but meaningless and invisible for human being. Generally, they lack explanatory power regardless of how accurate and complicated they are.
This letter jumps out of the shackle of existing works. 
To enhance interpretability, user behavior characteristics is employed in the user-behavior-based (UBB) NTP framework. 
Traffic is regarded as the superposition of several normal-distributed signals and the specific parameters are extracted from real-world traffic data. 
Numerical results show that it is a simple, efficient, accurate, and highly interpretable method to discover traffic patterns. 

Overall, the proposed method effectively utilizes the principal status of user behavior, and achieves the advantages as follows:
\begin{enumerate}
\item[1)]
Compared with existing methods, our method provides an interpretable NTP solution which is visual and comprehensible.
\item[2)]
Our method has a significant advantage in terms of computational efficiency.
\item[3)]
Our method establishes a correspondence between model parameters and user habits. The compatible expression of model parameters provides an opportunity to compare traffic patterns in different regions.
\end{enumerate}

The rest of the letter is structured as follows. In Section II, we detail the key points of the UBB NTP method and establish the mathematical model. In Section III, we simulate the UBB NTP method and benchmark methods, and perform a comparative analysis. Finally, we draw our conclusions and future plan in Section IV.

\section{The proposed UBB NTP method}

\noindent There are two approaches to gather user habits: one is from daily behavior, and the other is through real-world traffic data. Based on these habits, we build the mathematical model.

\begin{figure*}[!t]
\vspace{-0.6cm} 
\setlength{\abovecaptionskip}{-0.1cm} 
\setlength{\belowcaptionskip}{-3cm} 
\centering 
\includegraphics[width=5.75in]{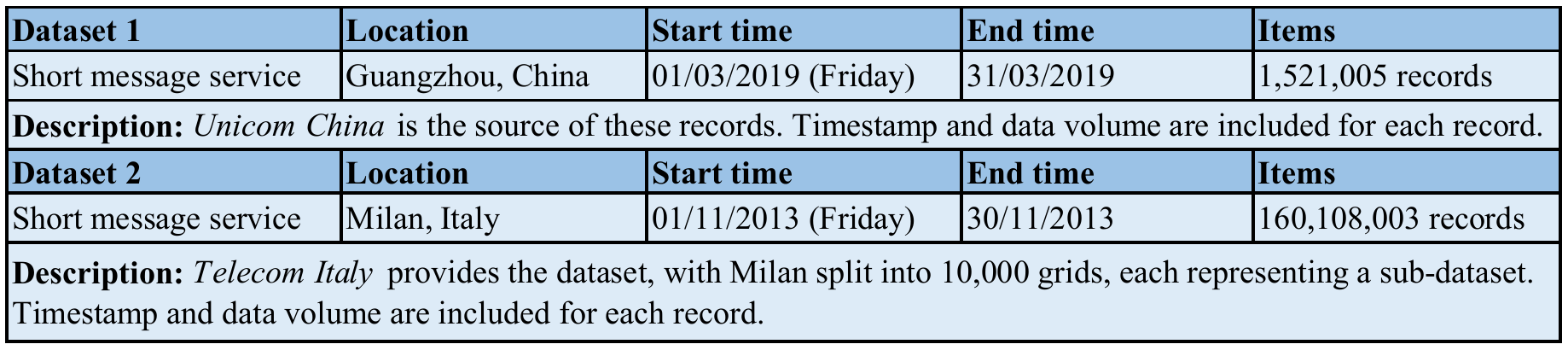} 
\caption{Description of the two adopted datasets.}
\label{fig1}
\end{figure*}

\subsection{Analysis for daily behavior}
\noindent Generally, people are accustomed to being active during the day and sleeping at night. 
In addition to sleeping hours, every day includes three main periods, i.e., morning, afternoon, and evening. Most people carry out their daily activities throughout these three periods. For example, every morning and afternoon on weekdays are generally working hours, whereas every evening is typically for recreation. These three periods thus correlate to the busiest times for network services. Hence, we divide the daily traffic into three traffic components, corresponding to the three time periods.

Furthermore, in the absence of prior information, we conceive that the peak time of each traffic component represents the time preference of users to use network.
Due to various subjective or objective factors, Some users' traffic demands deviate from the time preference.
The greater the magnitude of traffic deviation, the lower the probability of occurrence.
This assumption is well-founded because procrastination is prevalent in the population [12-14]. Statistically, the prevalence of procrastination is as high as 20-25$\%$ in the general population [13] and 15$\%$ of adults suffer from severe procrastination [14]. Correspondingly, there will be some users who like to finish their tasks in advance. 
In addition, outside the three main periods, the traffic caused by users' whims takes up a tiny percentage. For simplicity, all of the traffic outside the three main periods is regarded as an extension of these traffic components.

\begin{table}
\vspace{-0.5cm} 
\setlength{\abovecaptionskip}{0.1cm} 
\setlength{\belowcaptionskip}{-0.7cm} 
\begin{center}
\caption{The symbols corresponding to the 9 traffic components.}
\label{tab1}
\begin{tabular}{| c | c | c | c |}
\hline
 & Weekday & Saturday & Sunday\\
\hline
Morning & mw & msa & msu\\
\hline
Afternoon & aw & asa & asu\\
\hline
Evening & ew & esa & esu\\
\hline 
\end{tabular}
\end{center}
\end{table}


\begin{figure}[!t]
\vspace{-0.5cm} 
\setlength{\abovecaptionskip}{0.1cm} 
\setlength{\belowcaptionskip}{-0.5cm} 
\centering 
\includegraphics[width=3in]{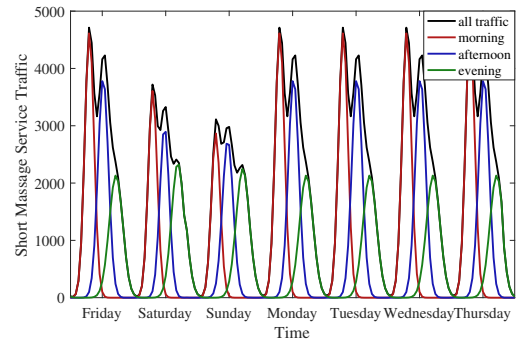}
\caption{Traffic distribution and the components in the morning, afternoon and evening.}
\label{fig1}
\end{figure}

\subsection{Analysis for traffic data}
\noindent 
This letter adopts two Short Message Service (SMS) datasets from Guangzhou, China, and Milan, Italy. The specifics of these datasets are shown in Fig. 1. Meanwhile, Dataset 1 is purchased and not publicly accessible, while Dataset 2 is available [15].

In both Datasets 1 and 2, the network traffic has a periodic variation during a week, which is consistent with the common sense that the week is a natural cycle of human activity. 
In addition, the traffic data on weekdays show a similar trend. It is mainly because the bulk of the urban populations leads a highly repetitive life on weekdays. The representative groups include students, teachers, enterprise employees, government officers, and so on. Hence, the workdays are characterized by the same traffic components.
Moreover, the employees in several occupations, like express industry, food service, etc., work seven days a week. Therefore, a portion of people maintains the pace of life on weekends. However, the traffic trends on Saturday and Sunday are different. We build dedicated models for Saturday and Sunday, respectively.

As shown in Fig. 2, we adopt three categories to represent the traffic on weekdays, Saturdays, and Sundays, respectively. Combined with the three main periods mentioned above, a total of nine traffic components are required to construct the UBB NTP method. The abbreviations in Table I represent these components.
The overall traffic is the superposition of all traffic components.
It is crucial to note that daily traffic could be represented by more or fewer components if there are sufficient social science data to back it up. It is a reflection of the scalability of the UBB NTP method.

\subsection{The mathematical model}
\noindent 
We adopt the SMS data in Guangzhou and Milan to extract parameters for mathematical model, respectively.
As shown in Fig. 2, the daily traffic curve is converted into three components. 
The red curve represents the traffic component distributed on the entire time axis with morning traffic as the main body.
Similarly, the blue and green ones correspond to the afternoon and evening components, respectively.

This letter considers a simple case. Assuming that the user behaviors that lead to traffic deviation obey the independent and identical distribution, 
then according to the central limit theorem, the prior distribution follows a Gaussian distribution.
Therefore, the morning traffic component on a certain workday is expressed as:

\begin{equation}
\setlength\abovedisplayskip{3pt}
\setlength\belowdisplayskip{3pt}
\label{deqn_ex1}
R_{\rm mw} =R_{{\rm mw},{\rm p}}{\exp\left( {- \frac{\left( {t - t_{{\rm mw},{\rm p}}} \right)^{\rm 2}}{{{\rm 2}\sigma}_{\rm mw}^{\rm 2}}} \right)},
\end{equation}

\noindent where $ {t}_{\rm mw,\rm p} $ is the time when each message is expected to be sent, $ {\sigma}_{\rm mw}^{\rm 2} $ is the variance and $ R_{\rm mw,\rm p} $ is the peak value. Similarly, we have:

\begin{figure}[!t]
\vspace{-0.5cm} 
\setlength{\abovecaptionskip}{0.1cm} 
\setlength{\belowcaptionskip}{-0.5cm} 
\centering 
\includegraphics[width=2.7in]{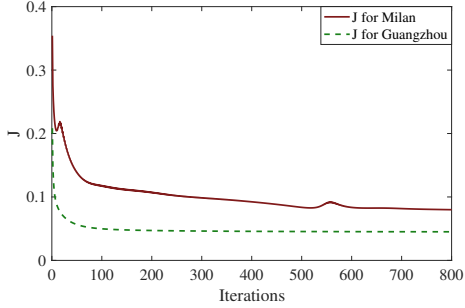} 
\caption{$J$ varies with the number of iterations.}
\label{fig1}
\end{figure}

\begin{equation}
\setlength\abovedisplayskip{3pt}
\setlength\belowdisplayskip{3pt}
\label{deqn_ex2}
R_{\rm c} = R_{\rm c,p}{\exp\left( {- \frac{\left( {t - t_{\rm c,p}} \right)^{\rm 2}}{{{\rm 2}\sigma}_{\rm c}^{\rm 2}}} \right)},
\end{equation}

\noindent in which $t$ is in a 24-hour format and $ {\rm c} = {\{ {\rm c_{1}},{\rm c_{2}},{\rm c_{3}} \}} $, represent weekday, Saturday and Sunday, respectively. Therefore, the hourly traffic at time $t$, the $k$th day of a week, can be represented as: 

\begin{equation}
\setlength\abovedisplayskip{3pt}
\setlength\belowdisplayskip{3pt}
\label{deqn_ex3}
R_{k}(t)={\sum_{ n_{\rm w}}\begin{bmatrix}
{\sum \limits_{n_{\rm d} = 1}\limits^{\rm 5}\sum\limits_{\rm c_{1}}^{}{R_{c_{1}}\left( {t + {\rm 24}\left( {n_{\rm d} - k} \right) + {{168n}_{\rm w}}} \right)}} \\
{+ {\sum\limits_{\rm c_{2}}^{}{{R_{\rm c_{2}}}\left( {t + {24(6 - k)} + {{168n}_{\rm w}}} \right)}}} \\
{+ {\sum\limits_{\rm c_{3}}^{}{{R_{\rm c_{3}}}\left( {t + {24(7 - k)} + {{168n}_{\rm w}}} \right)}}} \\
\end{bmatrix}},
\end{equation}

\noindent with the index of the day $k = [1,2,3,4,5,6,7]$, the index of the week number $ { n_{\rm w}} \in (- \infty, + \infty) $, the index of the weekday ${ n_{\rm d}}$, ${\rm c_{1}} = {\{ {\rm mw,aw,ew} \}}$, ${\rm c_{2}}= {\{{\rm msa,asa,esa}\}}$, and ${\rm c_{3}} = {\{{\rm msu,asu,esu}\}}$. After exchanging the sum order, we have:

\begin{equation}
\setlength\abovedisplayskip{3pt}
\setlength\belowdisplayskip{3pt}
\label{deqn_ex4}
\begin{matrix}
{R_{k}(t) = {\sum\limits_{\rm c_{1}}^{}{\sum\limits_{n_{\rm d} = 1}^{5}{\sum\limits_{n_{\rm w}}{{ R_{\rm c_{1}}}\left( {t + 24\left( {n_{\rm d} - k} \right) + {168n}_{\rm w}} \right)}}}}} \\
{+ {\sum\limits_{\rm c_{2}}^{}{\sum\limits_{ n_{\rm w}}{{ R_{\rm c_{2}}}\left( {t + 24\left( { 6 - k} \right) + {168n}_{\rm w}} \right)}}}} \\
{+ {\sum\limits_{\rm c_{3}}^{}{\sum\limits_{ n_{\rm w}}{{ R_{\rm c_{3}}}\left( {t + 24\left( { 7 - k} \right) + {168n}_{\rm w}} \right),}}}} \\
\end{matrix}
\end{equation}

\begin{equation}
\setlength\abovedisplayskip{3pt}
\setlength\belowdisplayskip{3pt}
\label{deqn_ex5}
\begin{small}
\begin{matrix}
{{R_{k}}(t) = {\sum\limits_{\rm c_{1}}^{}{{R_{\rm c_{1},p}}{\sum\limits_{n_{\rm d} = 1}^{5}{\sum\limits_{n_{\rm w}}{\exp\left( {- \frac{\left( {t + 24\left( {n_{\rm d} - k} \right) - t_{\rm c_{1},p} + {168n}_{\rm w}} \right)^{2}}{{2\sigma}_{\rm c_{1}}^{2}}} \right)}}}}}} \\
{+ {\sum\limits_{\rm c_{2}}^{}{{R_{\rm c_{2},p}}{\sum\limits_{n_{\rm w}}{\exp\left( {- \frac{\left( {t + 24\left( {6 - k} \right) - t_{\rm c_{2},p} + {168n}_{\rm w}} \right)^{2}}{{2\sigma}_{\rm c_{2}}^{2}}} \right)}}}}} \\
{+ {\sum\limits_{\rm c_{3}}^{}{{R_{\rm c_{3},p}}{\sum\limits_{n_{\rm w}}{\exp\left( {- \frac{\left( {t + 24\left( {7 - k} \right) - t_{\rm c_{3},p} + {168n}_{\rm w}} \right)^{2}}{{2\sigma}_{\rm c_{3}}^{2}}} \right)}}}}}, \\
\end{matrix}
\end{small}
\end{equation}
where ${ n_{\rm w}}$ is the domain of Gaussian signal.
For simplicity, we restrict the range of ${ n_{\rm w}}$ to [-1, 1]. Then, we have:

\begin{equation}
\setlength\abovedisplayskip{3pt}
\setlength\belowdisplayskip{3pt}
\label{deqn_ex6}
\begin{small}
\begin{matrix}
{{R_{k}}(t) \approx {\sum\limits_{\rm c_{1}}^{}{{R_{\rm c_{1},p}}{\sum\limits_{n_{\rm d} = 1}^{5}{\sum\limits_{n_{\rm w} = -1}^{+1}{\exp\left( {- \frac{\left( {t + 24\left( {n_{\rm d} - k} \right) - t_{\rm c_{1},p} + {168n}_{\rm w}} \right)^{2}}{{2\sigma}_{\rm c_{1}}^{2}}} \right)}}}}}} \\
{+ {\sum\limits_{\rm c_{2}}^{}{{R_{\rm c_{2},p}}{\sum\limits_{n_{\rm w} = -1}^{+1}{\exp\left( {- \frac{\left( {t + 24\left( {6 - k} \right) - t_{\rm c_{2},p} + {168n}_{\rm w}} \right)^{2}}{{2\sigma}_{\rm c_{2}}^{2}}} \right)}}}}} \\
{+ {\sum\limits_{\rm c_{3}}^{}{{R_{\rm c_{3},p}}{\sum\limits_{n_{\rm w} = -1}^{+1}{\exp\left( {- \frac{\left( {t + 24\left( {7 - k} \right) - t_{\rm c_{3},p} + {168n}_{\rm w}} \right)^{2}}{{2\sigma}_{\rm c_{3}}^{2}}} \right).}}}}} \\
\end{matrix}
\end{small}
\end{equation}

Thus, the parameter estimation problem becomes an optimization problem:

\begin{equation}
\setlength\abovedisplayskip{3pt}
\setlength\belowdisplayskip{3pt}
\label{deqn_ex7}
\underset{\begin{matrix}
{R_{\rm c,p},t_{\rm c,p},\sigma_{\rm c}^{2}} \\
{{\rm c} \in \left\{
{\begin{smallmatrix}
{\rm mw,aw,ew, }\\
{\rm msa,asa,esa, }\\
{\rm msu,asu,esu }\\
\end{smallmatrix}}
\right\}}
\end{matrix}}
{\rm minimise} J,
\end{equation}

\noindent where $J = \lVert {R_{k}}(t) - R_{\rm meas} \rVert^{2}$ and $R_{\rm meas}$ refers to the vector consisting of traffic measurements. There are many ways to solve this problem. This letter adopts a simple gradient descent method. As shown in Fig. 3, 
$J$ gradually decreases and converges as the number of iterations increases.

\section{Evaluation with real-world traffic data}
\noindent 
This section primarily compares the predictive accuracy and computational efficiency of the proposed UBB NTP method with benchmark methods. Predictive accuracy is measured by Mean Square Error (MSE), Root MSE (RMSE), Mean Absolute Error (MAE), and coefficient of determination (R2), which can be clearly defined by the following formulas:
\begin{equation}
\setlength\abovedisplayskip{3pt}
\setlength\belowdisplayskip{3pt}
\label{deqn_ex8}
{{\rm MSE}} = \frac{1}{n}\sum_{i=1}^n(y_i - \hat{y}_i)^2,
\end{equation}

\begin{equation}
\setlength\abovedisplayskip{3pt}
\setlength\belowdisplayskip{3pt}
\label{deqn_ex9}
{{\rm RMSE}} = \sqrt{{\rm MSE}},
\end{equation}

\begin{equation}
\setlength\abovedisplayskip{3pt}
\setlength\belowdisplayskip{3pt}
\label{deqn_ex10}
{{\rm MAE}} = \frac{1}{n}\sum_{i=1}^n\left|y_i - \hat{y}_i\right|,
\end{equation}

\begin{equation}
\setlength\abovedisplayskip{3pt}
\setlength\belowdisplayskip{3pt}
\label{deqn_ex11}
{{\rm R2}} = 1 - \frac{\sum_{i=1}^n(y_i - \hat{y}_i)^2}{\sum_{i=1}^n(y_i - \bar{y})^2},
\end{equation}
where $n$ represents the number of predicted samples, $y_i$ represents the actual value, $\hat{y}_i$ represents the predicted value, and $\bar{y}$ denotes the mean value of $y_i$.
Computational efficiency is represented by the elapsed time including training and prediction time. 

\begin{figure}[!t]
\vspace{-0.5cm} 
\setlength{\abovecaptionskip}{0.1cm} 
\setlength{\belowcaptionskip}{-0.5cm} 
\centering 
\includegraphics[width=2.7in]{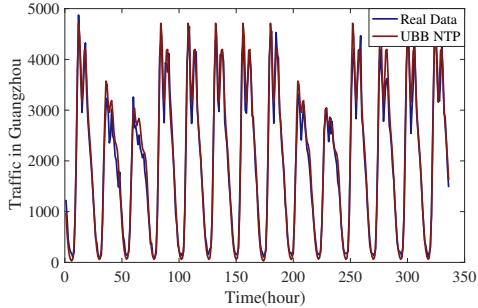} 
\caption{The prediction results of UBB NTP method for Guangzhou.}
\label{fig1}
\end{figure}

\begin{table}[htbp]
\centering
\begin{minipage}[t]{0.45\linewidth}
\centering
\caption{Parameters of UBB NTP method in Guangzhou.}
\label{tab:table1}
\begin{tabular}{| c | c | c | c |}
\hline
 & $R_{c,p}$ & $\sigma_{c}^{2}$ & $t_{c,p}$\\
\hline
mw & 4626 & 3.10 & 12.14\\
\hline
aw & 3839 & 3.91 & 17.35\\
\hline
ew & 2136 & 6.63 & 22.18\\
\hline
msa & 3612 & 2.89 & 12.09\\
\hline
asa & 2989 & 3.14 & 16.55\\
\hline
esa & 2356 & 5.78 & 21.60\\
\hline
msu & 2866 & 2.81 & 11.95\\
\hline
asu & 2759 & 4.13 & 16.47\\
\hline
esu & 2252 & 6.39 & 22.15\\
\hline 
\end{tabular}
\end{minipage}
\hspace{0.05\linewidth}
\begin{minipage}[t]{0.45\linewidth}
\centering
\caption{Parameters of UBB NTP method in Milan.}
\label{tab:table2}
\begin{tabular}{| c | c | c | c |}
\hline
 & $R_{c,p}$ & $\sigma_{c}^{2}$ & $t_{c,p}$\\
\hline
mw & 2942 & 0.88 & 9.45\\
\hline
aw & 6613 & 6.51 & 11.80\\
\hline
ew & 7327 & 13.91 & 18.72\\
\hline
msa & 3297 & 2.98 & 10.68\\
\hline
asa & 3684 & 9.92 & 13.87\\
\hline
esa & 4654 & 11.69 & 20.16\\
\hline
msu & 3720 & 5.29 & 11.43\\
\hline
asu & 3187 & 8.90 & 16.41\\
\hline
esu & 3843 & 8.59 & 21.36\\
\hline 
\end{tabular}
\end{minipage}
\end{table}

We choose a fully-connected Neural Network (NN) with five hidden layers and a standard recurrent NN which is an LSTM network with 3 hidden layers to represent ML-based methods. The ARMA model and ARIMA model are selected to represent statistics-based methods. According to the result of the Bayesian Information Criterion (BIC), the parameters are determined to be ARMA(4,2) for processing Dataset 1 and ARMA(3,1) for processing Dataset 2.
Following the same steps, the ARIMA models are set to ARIMA(2,1,3) and ARIMA(1,1,1) for Dataset 1 and Dataset 2, respectively.



\subsection{Performance of the UBB NTP method}

\noindent 
We use the SMS data of the first two weeks to build the proposed mathematical model and the rest to evaluate the performance.
As shown in Tables II and III, there are two parameter sets extracted from Dataset 1 and Dataset 2, respectively.
Then, we use the Guangzhou and Milan datasets to perform the NTP task.
The proposed UBB NTP method exhibits excellent performance in terms of prediction accuracy, as shown in Figs. 4 and 5. 
To some extent, it has verified that our assumptions regarding user behavior are realistic.

\subsection{Comparison with benchmark methods}

\noindent 
We first compare the performance of the UBB NTP method with all benchmark methods on the Guangzhou SMS data by the metrics of MSE, RMSE, MAE and R2.
As shown in Fig. 6, the UBB NTP method, which achieves the highest R2 and the lowest MSE  and RMSE, slightly outperforms the LSTM network and is superior to the statistics-based methods. Fig. 7 shows the prediction accuracy of each method on the Milan SMS data. As shown in Fig. 7, the UBB NTP method and the LSTM network have almost the identical MSE, RMSE and R2. Regardless of ARMA and ARIMA models, statistics-based methods do not perform very well in terms of accuracy. 
Since the UBB NTP method is designed with user behavior habits, it matches the practical wireless traffic demand very well.

Fig. 8 demonstrates the surprising superiority of the proposed UBB NTP method in terms of elapsed time. Take Dataset 1 as an example, our method completes both the training task and prediction step in only about 8.7 seconds. 
The efficiency is approximately 12 times the most efficient benchmark method, i.e., the ARMA model, and 28 times the most accurate benchmark method, i.e., the LSTM network. Although the multi-step mode saves some time for prediction, this time represents only a tiny fraction of the elapsed time. The key reason for the efficiency improvement is the reduced training time. Compared with benchmark methods, there is no need for offline training in the proposed UBB NTP method. 

Overall, the proposed UBB NTP method obtains the best overall performance in both Datasets 1 and 2, which means this method is well adapted to traffic data from different regions. 


\begin{figure}[!t]
\vspace{-0.5cm} 
\setlength{\abovecaptionskip}{0.1cm} 
\setlength{\belowcaptionskip}{-0.5cm} 
\centering 
\includegraphics[width=2.7in]{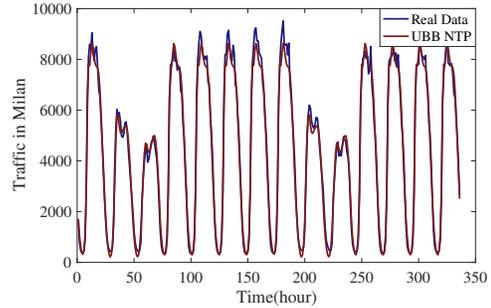} 
\caption{The prediction results of UBB NTP method for Milan.}
\label{fig1}
\end{figure}

\subsection{Analyses and discussion}
\noindent It is worth noting that the most significant advantage of the UBB NTP method is interpretability. It could be a link between traffic patterns and the field of social sciences. These parameters imply users' habits of using network traffic.
In Tables II and III, $ {R}_{\rm c,\rm p} $ denotes the peak of the traffic component, while $ {t}_{\rm c,\rm p} $ denotes the corresponding time in a 24-hour format, measured in hours.
Take the parameter set of Guangzhou as an example. 
According to the 3-sigma rule in the normal distribution, 68\% of morning traffic occurs between 10:15 am and 1:55 pm, 68\% of traffic demand in every afternoon occurs between 2:30 pm and 7:20 pm, and 68\% of traffic demand in every evening occurs from 7:10 pm to 0:45 am the next day.
On both weekdays and weekends, the morning has the highest traffic demand, followed by the afternoon, and the evening has the lowest.

Moreover, $\sigma_{c}^{2}$ indicates the magnitude of traffic deviation; the higher the $\sigma_{c}^{2}$, the more dispersed the distribution. In both Tables II and III, we can observe that $\sigma_{c}^{2}$ at night are higher than those during the day. It mainly caused by the changes in users' state. As the day progresses, users gradually change from a working state to a leisure state. After that, users generate traffic whenever and wherever they want just for their individual needs, such as chatting, ordering take-out, etc.
The distribution is, therefore, more dispersed at night. As shown in Table III, $\sigma_{ew}^{2}$ and $R_{ew,p}$ means a large number of users are active during the period. What's more, the leisure users dominate.

By comparing Milan and Guangzhou, the traffic curves in the two cities in Figs. 4 and 5 are comparable. To some extent, it verifies that urban users have similar living habits. Furthermore, both Milan and Guangzhou have highly developed tertiary industries [16-17], which could be one of the main reasons for the similar traffic curves. 
On the other hand, Tables II and III demonstrate that users in Guangzhou and Milan have various preferences. 
The distribution of peak times of the three traffic components is more even in Guangzhou, and they are all spaced about five hours apart. While in Milan, the first two traffic components are close in time and farther apart from the one representing evening.

\begin{figure}[!t]
\vspace{-0.8cm} 
\setlength{\abovecaptionskip}{0.1cm} 
\setlength{\belowcaptionskip}{-0.5cm} 
\centering 
\includegraphics[width=2.7in]{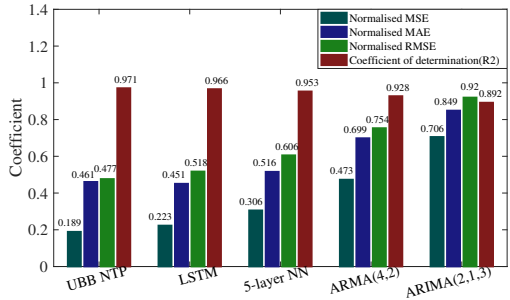} 
\caption{The prediction performance of the proposed UBB NTP method and the benchmark methods for Guangzhou SMS dataset}
\label{fig1}
\end{figure}

\begin{figure}[!t]
\vspace{-0.45cm} 
\setlength{\abovecaptionskip}{0.1cm} 
\setlength{\belowcaptionskip}{-0.5cm} 
\centering 
\includegraphics[width=2.7in]{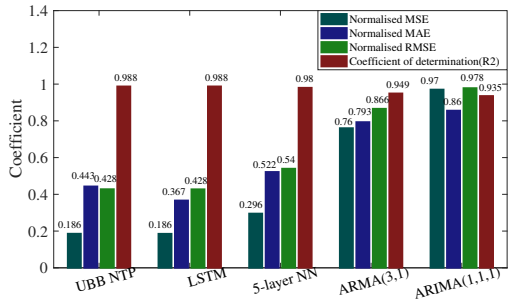} 
\caption{The prediction performance of the proposed UBB NTP method and the benchmark methods for Milan SMS dataset}
\label{fig1}
\end{figure}

\section{Conclusion}
\noindent 
In this letter, we introduce a novel UBB NTP method that exhibits higher overall performance when compared to existing ML-based and statistics-based methods. The method's parameters are concise, possess practical significance, and provide an interpretable NTP solution.
Furthermore, the standardized parameter set enables comparing traffic patterns across different regions. Hence, the proposed UBB NTP method may be considered a promising aspect in the combination of communication and social science.

\begin{figure}[!t]
\vspace{-0.8cm} 
\setlength{\abovecaptionskip}{0.1cm} 
\setlength{\belowcaptionskip}{-0.5cm} 
\centering 
\includegraphics[width=2.7in]{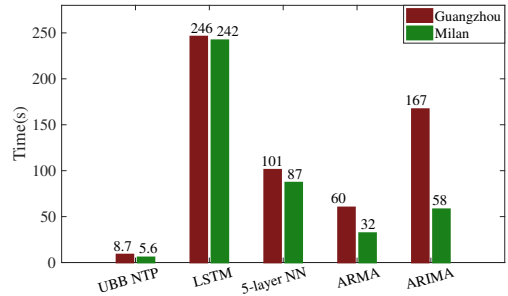} 
\caption{The elapsed time of the proposed UBB NTP method and the benchmark methods.}
\label{fig1}
\end{figure}

Our future research will explore similarities and differences among different types of traffic data. To accurately predict long-term traffic trends, we will develop a seasonal model that complements the UBB NTP model. We will also focus on detecting anomalous traffic caused by infrequent events by analyzing patterns and changes. Additionally, we will employ meta-learning to speed up and optimize parameter selection and further improve prediction accuracy and computational efficiency of the proposed method.

\end{document}